\documentclass[aps,prb,twocolumn,superscriptaddress,letterpaper,10pt,floatfix]{revtex4-2}
\usepackage{amsmath,amssymb,bm,graphicx,xcolor}
\usepackage[colorlinks=true,citecolor=blue!60!black,urlcolor=blue!60!black,linkcolor=blue!60!black]{hyperref}
\newcommand{\qpar}{q_\parallel}

\begin{document}

\title{Uniaxial strain induces $d$-wave phonon chirality in ferroelectric BaTiO$_3$}

\author{Juan M. Florez}
\email{jflorezu.docente@ucen.cl}
\affiliation{Facultad de Ingenier\'ia y Arquitectura, Universidad Central de Chile, Santiago, Chile}
\author{Eric Su\'arez Morell}
\email{eric.suarez@usm.cl}
\affiliation{Grupo de Simulaciones, Departamento de F\'isica, Universidad T\'ecnica Federico Santa Mar\'ia, Valpara\'iso, Chile}

\date{\today}

\begin{abstract}
Circularly dichroic resonant inelastic x-ray scattering has revealed electrically switchable $g$-wave phonon chirality in freestanding BaTiO$_3$ membranes. Here, we show from first principles that uniaxial strain changes the angular class of this texture. Lowering $C_{4v}$ to $C_{2v}$ adds a $d$-wave helicity that is bilinear in the orthorhombic strain $\eta$ and the polarization, with nodal lines along the stress axis. For the 38~meV branches at $\eta=1\%$, we find that (i) along the diagonal, where the helicity vanishes by symmetry without strain, it reaches about 60\% of the unstrained maximum, and (ii) at the measured wave vectors off the mirror it is reduced by more than half. With PBEsol and PBE+D3 force constants, the $d$-wave component grows linearly in $\eta$ and equals the $g$-wave one at $\eta^*\approx1.3\text{--}1.4\%$. Resolved by branch, the response of strain-split $E$ doublets is non-analytic in $\eta$, as captured by a two-band model. Strain therefore provides a second knob, besides the electric field, that controls the symmetry and the sign of phonon chirality in ferroelectric membranes.
\end{abstract}

\maketitle

\section{Introduction}
Phonons with circular atomic motion carry angular momentum, which couples to spins, magnetization and heat~\cite{ZhangNiu2014,Juraschek2019,Zhu2018,Juraschek2025}. In achiral crystals the helicity cannot survive a momentum-space average, but it can form higher-harmonic helicity textures whose pattern is fixed by the crystal point group, as established by symmetry-based classifications~\cite{Coh2023,Yang2026,Zhang2026}. Following the detection of chiral phonons in quartz by circularly dichroic resonant inelastic x-ray scattering (RIXS)~\cite{Ueda2023}, such textures have been observed in polar LiNbO$_3$~\cite{Ueda2025} and coupled to ferroelectricity in a molecular crystal~\cite{Han2026}. Recently, Grimes \emph{et al.}\ used this technique to resolve a $g$-wave helicity texture in freestanding BaTiO$_3$ membranes and to reverse it with an electric field~\cite{Grimes2026}, establishing ferroelectric membranes as a platform for switchable phonon chirality. This result builds on the earlier prediction by Moseni \emph{et al.}\ that the phonon angular momentum of tetragonal BaTiO$_3$ lies mostly perpendicular to the polarization and follows its reversal~\cite{Moseni2022}.

\begin{figure*}[t]
\includegraphics[width=\textwidth]{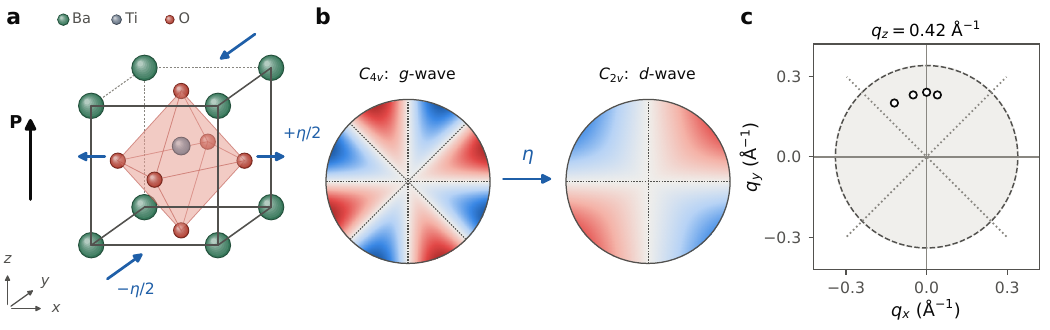}
\caption{(a) Polar P4mm cell (tetragonality and polar displacements exaggerated) under orthorhombic strain $\eta=\varepsilon_{xx}-\varepsilon_{yy}$, which lowers $C_{4v}$ to $C_{2v}$. (b) Schematic helicity patterns: $g$-wave ($\propto\sin4\phi$) and the strain-induced $d$-wave component ($\propto\sin2\phi$); dotted lines are the nodal lines fixed by the mirrors. (c) In-plane momentum space at $q_z=0.42$~\AA$^{-1}$: the shaded disk is the reach of O $K$-edge RIXS, circles mark the four measured wave vectors of Ref.~\cite{Grimes2026}, and solid (dotted) lines are the mirrors kept (lost) under $\eta$.}
\label{fig:concept}
\end{figure*}

Polarization reversal changes the sign of the texture while preserving its fourfold pattern, which raises the question of whether a second, independent knob can change the angular class itself. $d$-wave textures of phonon angular momentum have been predicted in altermagnets, where they originate from broken time-reversal symmetry~\cite{Chatzichrysafis2026,Bendin2026}, Pols and Spaldin have demonstrated that symmetry lowering at surfaces makes phonons chiral~\cite{Pols2026}, and $d$-wave textures of a different quantity, the sublattice polarization, have been proposed for nonmagnetic crystals~\cite{You2026}. In a nonmagnetic ferroelectric, the natural candidate is strain, all the more so because freestanding oxide membranes sustain uniaxial strains of several percent~\cite{Dong2019,Xu2020,Hong2020}. Here, we take the system and the wave vectors of Ref.~\cite{Grimes2026} as our starting point and show from first principles that strain adds a $d$-wave helicity to the measured $g$-wave texture, i.e., the texture that the electric field switches can also be reshaped.

\section{Symmetry and scaling}
For a branch $\nu$ with mass-weighted eigenvector $\mathbf e_\nu(\mathbf q)$, the phonon angular momentum is $\bm\ell_\nu=\sum_\kappa\mathbf e^\dagger_{\kappa\nu}\mathbf M\mathbf e_{\kappa\nu}$ in units of $\hbar$, with $(M_\alpha)_{\beta\gamma}=-i\epsilon_{\alpha\beta\gamma}$~\cite{ZhangNiu2014}, and the helicity is $h_\nu=\bm\ell_\nu\cdot\hat{\mathbf q}$. Time reversal gives $\bm\ell_\nu(-\mathbf q)=-\bm\ell_\nu(\mathbf q)$, so that $h_\nu(-\mathbf q)=h_\nu(\mathbf q)$, and $h$ changes sign under every mirror. In $C_{4v}$ with polar axis $z$, $h$ therefore transforms as $A_2$. Since inversion and time reversal together force $h\equiv0$ in a centrosymmetric crystal, every allowed coefficient is odd in $P$. The orthorhombic strain $\eta=\varepsilon_{xx}-\varepsilon_{yy}$ transforms as $B_1$ and the shear $\gamma=2\varepsilon_{xy}$ as $B_2$. Because $B_1\otimes B_2=A_2$, the invariant expansion at fixed $q_z$ reads
\begin{equation}
h=A\,q_xq_y(q_x^2-q_y^2)+B\,\eta\,q_xq_y+B'\gamma\,(q_x^2-q_y^2)+\dots,
\label{eq:h}
\end{equation}
with $A$, $B$ and $B'$ odd in $P$, even in $q_z$, and independent of each other. Stress along $[100]$ thus adds a $\sin2\phi$ texture that lifts the diagonal nodes, while the axial nodes remain protected by the surviving mirrors [Fig.~\ref{fig:concept}(b,c)]. Stress along $[110]$ adds a $\cos2\phi$ texture with the opposite node pattern. The $d$-wave term is bilinear in polarization and strain, its nodal lines follow the stress axis, and it is odd in $\eta$, so that tensile and compressive stress give the same magnitude with opposite signs. For a uniaxial stress with Poisson ratio $\nu$, $\eta=(1+\nu)\varepsilon$; the isotropic part $\varepsilon_{xx}+\varepsilon_{yy}$ preserves $C_{4v}$ and only renormalizes $A$.

On a ring of radius $\qpar$, Eq.~\eqref{eq:h} gives the Fourier amplitudes $b_4=A\qpar^4/4$ and $b_2=B\eta\qpar^2/2$. The $d$-wave component dominates for
\begin{equation}
\eta>\eta^*(\qpar)=\left|\frac{A}{2B}\right|\qpar^2 .
\label{eq:etastar}
\end{equation}
At the O $K$ edge (531~eV) the transferred momentum is limited to $|\mathbf Q|\le0.54$~\AA$^{-1}$, and the measured points lie at $\qpar\approx0.23$~\AA$^{-1}$ and $q_z=0.42$~\AA$^{-1}$~\cite{Grimes2026}. Equation~\eqref{eq:etastar} shows that $\eta^*$ grows with $\qpar$, so that the measured points, at the edge of the RIXS reach, are the least favorable; a quantitative value requires the first-principles amplitudes [Fig.~\ref{fig:pred}(a)].

\begin{figure*}[t]
\includegraphics[width=\textwidth]{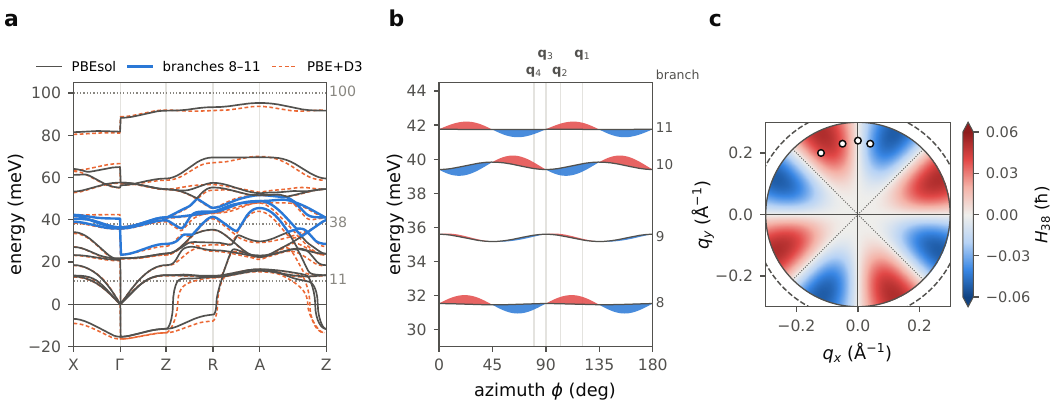}
\caption{Unstrained P4mm BaTiO$_3$ (PBEsol, $3\times3\times3$ interpolation). (a) Phonon dispersion with PBEsol (solid) and PBE+D3 (dashed, $2\times2\times2$) force constants, including LO--TO splitting; the PBEsol branches 8--11 (in energy order), which form the 38~meV group, are shown in blue. Imaginary frequencies are shown as negative, and dotted lines mark the 11, 38 and 100~meV modes of Ref.~\cite{Grimes2026}. (b) Branches 8--11 on the measured ring ($\qpar=0.23$, $q_z=0.42$~\AA$^{-1}$) versus azimuth: each line is the mode energy, and the filled excursion is proportional to the mode helicity (red positive, blue negative; 12~meV per $\hbar$). Every branch changes sign at the mirrors, $\phi=0^\circ$, $45^\circ$, $90^\circ$, $135^\circ$. Vertical guides mark the azimuths of the measured wave vectors $\mathbf q_{1\text{--}4}$. (c) Group helicity $H_{38}(q_x,q_y)$ at $q_z=0.42$~\AA$^{-1}$, a $g$-wave texture. The dashed circle is the RIXS reach, open circles are the measured wave vectors, and solid (dotted) lines are the axial (diagonal) mirrors.}
\label{fig:bench}
\end{figure*}

\section{Strain-split doublets}
On the $\Gamma$--Z line the 15 branches decompose into $4A_1+B_1+5E$. Most branches near the measured $\mathbf q$ therefore derive from $E$ doublets, where Eq.~\eqref{eq:h} is not uniform in $\qpar$. We capture this regime with a two-band model in the basis of the doublet partners, $D(\mathbf q)=\omega_0^2+\mathbf d(\mathbf q)\cdot\bm\sigma$, with
\begin{align}
d_z&=b\,(q_x^2-q_y^2)+\kappa\eta,\qquad d_x=2c\,q_xq_y,\nonumber\\
d_y&=q_z\big[g\,q_xq_y(q_x^2-q_y^2)+\beta\eta\,q_xq_y\big].
\label{eq:model}
\end{align}
At fixed $q_z\neq0$ the dipolar (LO--TO) term is analytic in $\qpar$ and isotropic in the plane, so it only renormalizes $b$ and $c$. Time reversal requires the antisymmetric ($\sigma_y$) coefficient to be odd in $\mathbf q$, and for the two eigenmodes $\ell_z=\mp d_y/|\mathbf d|$ exactly. Two consequences follow. At $\eta=0$ the doublet gap closes as $\qpar^2$, so the $g$-wave helicity of these branches scales as $\qpar^2$ rather than $\qpar^4$. For $\qpar\ll q^*=\sqrt{\kappa|\eta|/b}$ the strain gap dominates and
\begin{equation}
h_d\simeq\mathrm{sgn}(\eta)\,\frac{\beta\,q_z^2}{2\kappa|\mathbf q|}\,\qpar^2\sin2\phi ,
\label{eq:sat}
\end{equation}
which is independent of $|\eta|$, whereas the $g$-wave component decays as $1/\eta$. The branch-resolved response is therefore non-analytic in $\eta$ ($\propto\mathrm{sgn}\,\eta$) and saturates for $\qpar\ll q^*$. Since the two partners carry opposite $\ell_z$, their sum cancels to this order: the saturation is a property of the individual split branches, which requires an energy resolution of the order of $\kappa\eta$ to be observed, whereas the energy-integrated signal follows Eq.~\eqref{eq:h}. Its onset is set by the splitting $\kappa\eta$ of the $E$ modes, which Raman scattering can measure independently.

Table~\ref{tab:gamma} gives $\kappa$ from first principles. The soft $E$ mode, harmonically unstable at $\Gamma$ as usual for tetragonal BaTiO$_3$ at $T=0$, splits by about 70~cm$^{-1}$ per percent; at $\eta=2\%$ one partner becomes real, while the partner polarized along the tensile axis softens further. The stiffer doublets split by 1--8~cm$^{-1}$ per percent.

\begin{table}[b]
\caption{$\Gamma$-point $E$ doublets (cm$^{-1}$, PBEsol, DFPT without LO--TO splitting) under orthorhombic strain $\eta$. Negative values denote imaginary frequencies.}
\label{tab:gamma}
\footnotesize\setlength{\tabcolsep}{2.5pt}
\begin{ruledtabular}
\begin{tabular}{lcccc}
$E$ & $\eta=0$ & $0.5\%$ & $1\%$ & $2\%$\\\hline
soft & $-122$ & $-101/-135$ & $-74/-153$ & $63/-179$\\
176 & 175.7 & 176.4/175.4 & 175.4/175.3 & 175.3/174.9\\
294 & 294.5 & 295.3/293.5 & 296.3/292.3 & 298.3/289.8\\
465 & 464.9 & 466.7/463.2 & 468.8/460.6 & 473.3/456.7\\
\end{tabular}
\end{ruledtabular}
\end{table}

\section{First-principles calculations}
We compute total energies and forces with VASP 6.4.2~\cite{Kresse1996,Kresse1999} in the projector-augmented-wave formalism, treating Ba $5s5p6s$ and Ti $3s3p3d4s$ as valence. We compare two functionals: PBEsol~\cite{Perdew2008}, the standard for BaTiO$_3$ lattice dynamics~\cite{Moseni2022}, and PBE with D3(BJ) dispersion~\cite{Perdew1996,Grimme2010,Grimme2011}, the functional used to model the experiment~\cite{Grimes2026}. Electronic states are converged to $10^{-8}$~eV with Gaussian smearing of 0.02~eV and aspherical PAW corrections. Five-atom calculations use 800~eV and a $\Gamma$-centered $8^3$ mesh. The relaxed P4mm cells have $a=3.968$, $c=4.070$~\AA\ ($c/a=1.026$, PBEsol) and $a=3.986$, $c=4.116$~\AA\ ($c/a=1.033$, PBE+D3), with residual forces below $5\times10^{-4}$~eV/\AA\ and stresses below 0.3~kB; experiment gives 4.00 and 4.03~\AA. The PBE+D3 cell reproduces the published ABINIT values of 3.99 and 4.11~\AA~\cite{Grimes2026}. Strained cells are obtained from the relaxed $a_0$ with $\varepsilon_{xx}=-\varepsilon_{yy}=\eta/2$ (or $\varepsilon_{xy}=\gamma/2$), with $c$ from a seven-point energy scan and relaxed ions.

\begin{figure*}[t]
\includegraphics[width=\textwidth]{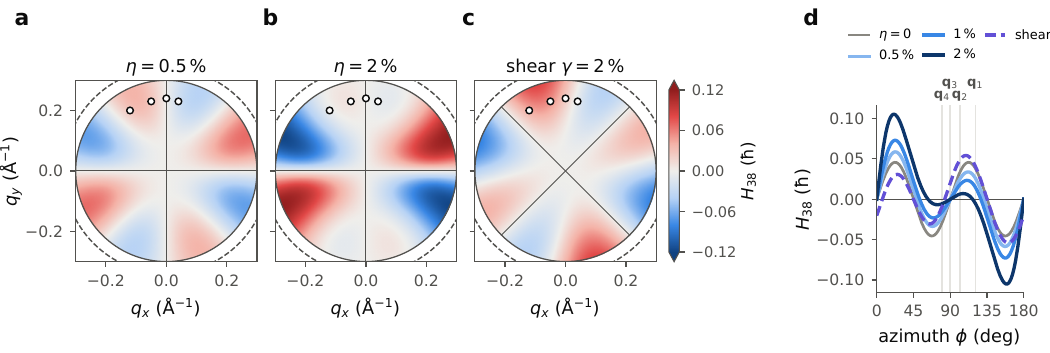}
\caption{Strain control (PBEsol, $3\times3\times3$ interpolation unless stated). (a--c) Group helicity $H_{38}$ at $q_z=0.42$~\AA$^{-1}$ for $\eta=0.5\%$ ($3\times3\times3$, interpolated), $\eta=2\%$ ($3\times3\times3$) and $2\%$ shear ($2\times2\times2$), on a common color scale; solid lines are the mirrors kept under each strain. (d) $H_{38}$ on the measured ring versus azimuth for $\eta=0$, 0.5, 1 and 2\% ($3\times3\times3$) and for 2\% shear ($2\times2\times2$); vertical lines mark the azimuths of $\mathbf q_{1\text{--}4}$.}
\label{fig:strain}
\end{figure*}

Force constants follow from finite displacements of $\pm0.01$~\AA\ in $2\times2\times2$ supercells (40 atoms; 8 displacements in P4mm, 10 in Pmm2 and Cmm2) with Phonopy~\cite{Togo2015,Togo2023}, at 650~eV and a $\Gamma$-centered $3^3$ mesh. A $4^3$ $k$ mesh in the supercell changes $b_4$ by 0.3\%, an 800~eV cutoff changes the forces by at most $2.4\times10^{-3}$~eV/\AA, and a displacement of 0.02~\AA\ changes $b_4$, $b_2$ and $H_{38}(\mathbf q_{1\text{--}4})$ by less than 0.6\%. The helicity texture is, however, sensitive to long-range force constants: in a $3\times3\times3$ supercell (135 atoms, $\Gamma$-centered $2^3$ mesh) the mode energies of the 38~meV group change by less than 1~meV, but its $g$-wave amplitude increases by 16\% at the measured $\qpar$. The strain response is even more sensitive: at $\eta=2\%$ the $d$-wave amplitude grows by 61\%, so that the crossover strain introduced below decreases from 2.2\% ($2\times2\times2$) to 1.6\% ($3\times3\times3$) (Table~\ref{tab:sc}). As checks free of Fourier interpolation, we use nondiagonal supercells~\cite{LloydWilliams2015} with converged $k$ meshes in compact cells. A 50-atom cell commensurate with $\mathbf q=(0.1,0.1,0.3)$ (reduced units; $\qpar=0.224$~\AA$^{-1}$, $\phi\approx45^\circ$, $q_z=0.46$~\AA$^{-1}$) gives $H_{38}=0.047\,\hbar$ at $\eta=2\%$, with the $3\times3\times3$ value 3\% below it. A 75-atom cell commensurate with $\mathbf q=(2/15,1/15,4/15)$, on the measured ring ($\qpar=0.236$~\AA$^{-1}$, $\phi=26.6^\circ$, $q_z=0.41$~\AA$^{-1}$), gives $H_{38}=0.036\,\hbar$ at $\eta=0$, with the $3\times3\times3$ value 24\% and the $2\times2\times2$ value 8\% above it; at this azimuth $H_{38}$ is set by $b_4$ ($b_8<0.001\,\hbar$). At the wave vectors they share with the diagonal supercells, both nondiagonal sets reproduce the mode energies to 0.02~meV, and at $\Gamma$ they reproduce the DFPT frequencies. The groups below 26~meV, which hybridize with the soft mode, are not converged at this level and are not used in the predictions. We report PBEsol amplitudes from $3\times3\times3$ supercells, anchored to these checks where stated; anchoring multiplies the unstrained helicity by 0.81 and the strain-induced change by 1.03. At $\eta=0.5$ and 1\% we obtain them from the force constants at 0 and 2\%, scaling the change that is odd under the $90^\circ$ rotation (99\% of it) linearly in $\eta$ and the even part quadratically, with the relaxed cell and Born charges of each strain; at $2\times2\times2$, where all strains were computed directly, this reproduces $b_2$, $H_{38}(\mathbf q_{1\text{--}4})$ and the branch helicities within 2\%. The PBE+D3 $\eta^*$ is taken as 11\% above the PBEsol value, the ratio found at $2\times2\times2$. Dipole--dipole interactions are included through the Gonze--Lee correction~\cite{Gonze1997} with Born charges and $\varepsilon^\infty$ from density-functional perturbation theory (DFPT). In PBEsol, $Z^*_{\rm Ti}=7.06$ ($\perp c$) and $\varepsilon^\infty_{\perp}=6.46$, the anomalous charges characteristic of BaTiO$_3$~\cite{Ghosez1998}. Born charges and $\varepsilon^\infty$ are recomputed for every strained cell. Since D3 does not act on the electrons, PBE+D3 Born charges are computed with PBE at the PBE+D3 geometry. The $\Gamma$-point frequencies of the finite-displacement force constants agree with DFPT to within 2\%.

\begin{table}[t]
\caption{Supercell dependence (PBEsol) of the $g$-wave ($b_4$, $\eta=0$) and $d$-wave ($b_2$, $\eta=2\%$) amplitudes, in $\hbar$, and of the crossover strain $\eta^*$ of the 38~meV group at two values of $\qpar$ (\AA$^{-1}$). The last row anchors the $3\times3\times3$ values to the nondiagonal checks (factors 0.81 for $b_4$ and 1.03 for $b_2$). Unless stated, PBEsol values in the text are $3\times3\times3$ ($\eta=0.5$ and 1\%: interpolated), and PBE+D3 values are $2\times2\times2$.}
\label{tab:sc}
\footnotesize\setlength{\tabcolsep}{3pt}
\begin{ruledtabular}
\begin{tabular}{llcccccc}
 & & \multicolumn{3}{c}{$2\times2\times2$} & \multicolumn{3}{c}{$3\times3\times3$}\\
group & $\qpar$ & $b_4$ & $b_2$ & $\eta^*$ & $b_4$ & $b_2$ & $\eta^*$\\\hline
$G_{38}$ & 0.14 & 0.012 & 0.033 & 0.75\% & 0.016 & 0.045 & 0.71\%\\
$G_{38}$ & 0.23 & 0.039 & 0.035 & 2.23\% & 0.046 & 0.057 & 1.60\%\\
anchored & 0.23 & & & & 0.037 & 0.059 & 1.26\%\\
\end{tabular}
\end{ruledtabular}
\end{table}

We evaluate $\bm\ell_\nu$ and $h_\nu$ at Cartesian wave vectors on rings of radius $\qpar=0.04$--$0.30$~\AA$^{-1}$ at $q_z=0.42$~\AA$^{-1}$ and at the four measured points of Ref.~\cite{Grimes2026}. Because bands disperse across any fixed energy window, we sum the helicity over groups of branches in energy order, $H_G(\mathbf q)=\sum_{\nu\in G}h_\nu$, which is invariant under rotations within degenerate subspaces as long as neighboring groups do not touch. The 38~meV feature corresponds to branches 8--11 ($G_{38}$; 32.6--41.4~meV at $\mathbf q_1$), separated from the adjacent groups by at least 3.7~meV everywhere. The two lowest branches, which contain the soft mode, are excluded. The sum of $h_\nu$ over all branches, symmetry-forbidden harmonics and $H_G(-\mathbf q)-H_G(\mathbf q)$ vanish to numerical precision.

\begin{figure*}[t]
\includegraphics[width=\textwidth]{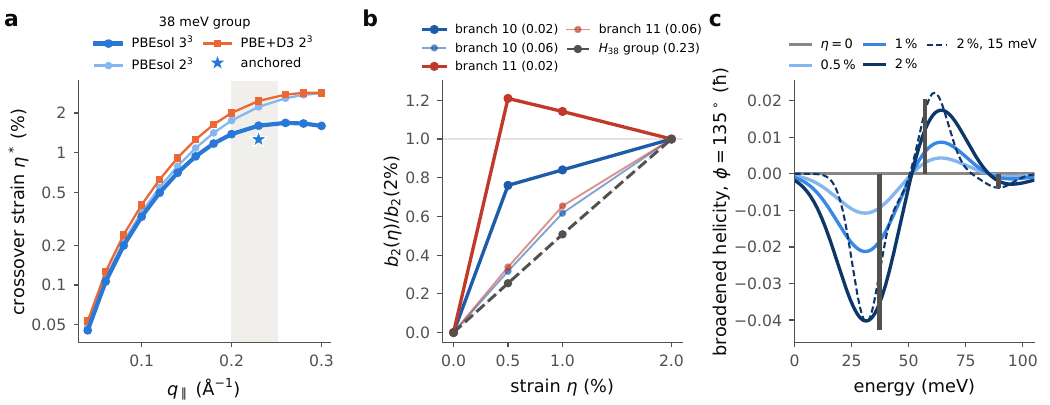}
\caption{Predictions. (a) Crossover strain $\eta^*$ of the 38~meV group versus $\qpar$: PBEsol in $3\times3\times3$ supercells (from $\eta=2\%$, thick; star: anchored to the nondiagonal checks), PBEsol and PBE+D3 in $2\times2\times2$ supercells (from $\eta=1\%$); the shaded band marks the measured $\qpar$. (b) $d$-wave amplitude versus strain, normalized to $\eta=2\%$ ($3\times3\times3$; $\eta=0.5$ and 1\% interpolated): below $q^*$ ($\qpar=0.02$~\AA$^{-1}$) the partners of the 39~meV doublet depend only weakly on $\eta$ [Eq.~\eqref{eq:sat}], above it (0.06~\AA$^{-1}$, faint) and for the group sum $H_{38}$ at the measured $\qpar$ (dashed) the response is linear (dotted line). Values in parentheses are $\qpar$ in \AA$^{-1}$. (c) Partial spectrum of the converged branches along the diagonal, $\sum_{\nu=8}^{15}h_\nu G(E-\hbar\omega_\nu)$ at $\qpar=0.23$~\AA$^{-1}$, $\phi=135^\circ$, $q_z=0.42$~\AA$^{-1}$, with a Gaussian $G$ of 30~meV FWHM, the resolution of Ref.~\cite{Grimes2026}, for $\eta=0$, 0.5, 1 and 2\% (PBEsol, $3\times3\times3$; dashed: $\eta=2\%$ at 15~meV FWHM); it vanishes by symmetry at $\eta=0$. Vertical bars are the group helicities $H_{38}$, $H_{57}$ and $H_{89}$ ($3\times3\times3$, $\eta=2\%$) at the mean energy of each group; branches below 26~meV, which hybridize with the soft mode, are omitted (see text).}
\label{fig:pred}
\end{figure*}

\section{Results}
\textit{Benchmark.} At $\mathbf q_1$ the PBEsol branches lie at 12.4--13.5, 32.6--41.4 and 88.6~meV, consistent with the modes at 11, 38 and 100~meV to which Ref.~\cite{Grimes2026} attributes the observed dichroic contrast. The summed helicity of the 38~meV group is $+0.039$, $+0.035$, $0$ and $-0.029\,\hbar$ at $\mathbf q_{1\text{--}4}$ ($3\times3\times3$): it vanishes on the mirror ($\mathbf q_3$) and changes sign across it, as the measured dichroism does. Resolved by branch, every mode of the group changes sign at the four mirrors on the measured ring [Fig.~\ref{fig:bench}(b)], and the group sum is a clean $g$-wave texture [Fig.~\ref{fig:bench}(c)], with $b_4=0.046\,\hbar$ at $\qpar=0.23$~\AA$^{-1}$ ($3\times3\times3$; $0.037\,\hbar$ anchored to the nondiagonal check) that grows as $\qpar^{3.4}$ at small $\qpar$, close to the leading $\qpar^4$ of Eq.~\eqref{eq:h}. PBE+D3 gives the same picture [Fig.~\ref{fig:bench}(a)], with mode energies within 0.4~meV and $b_4$ within 3\% at the measured $\qpar$ ($2\times2\times2$), although the two functionals differ in $c/a$.

\textit{Strain.} Orthorhombic strain lifts the diagonal nodes and adds a $\sin2\phi$ component [Fig.~\ref{fig:strain}(a,b)], while shear adds a $\cos2\phi$ component and rotates the nodal lines by $45^\circ$ [Fig.~\ref{fig:strain}(c)]. On the measured ring, $H_{38}$ develops a growing $\sin2\phi$ component [Fig.~\ref{fig:strain}(d)]. The $d$-wave amplitude rises as $\qpar^{1.7}$ at the smallest momenta, close to the $\qpar^2$ of Eq.~\eqref{eq:h}, but grows more slowly above $\qpar\approx0.15$~\AA$^{-1}$ and reaches $b_2=0.057\,\hbar$ at the measured $\qpar$ ($\eta=2\%$), and therefore we use the computed amplitudes directly. It is linear in $\eta$: $b_2(2\%)/b_2(1\%)=1.99$ (PBEsol) and 2.01 (PBE+D3), both computed directly at $2\times2\times2$, where PBE+D3 lies 11\% below PBEsol with the same $\qpar$ dependence.

\textit{Doublet saturation.} Resolved by branch, the response follows Eq.~\eqref{eq:sat}. For the 39~meV doublet (branches 10 and 11), $\kappa\eta=0.18$~meV at $\eta=0.5\%$ and the zero-strain splitting gives $b\approx150$~meV\,\AA$^2$, so that $q^*\approx0.035$~\AA$^{-1}$. Below $q^*$, at $\qpar=0.02$~\AA$^{-1}$, the two partners carry $b_2\approx-0.004$ and $+0.003\,\hbar$, the same at $\eta=0.5$, 1 and 2\% within 30\% and growing as $\qpar^2$ [Fig.~\ref{fig:pred}(b)]; above $q^*$, at $\qpar=0.06$~\AA$^{-1}$, they grow nearly linearly with $\eta$ ($-0.007$, $-0.014$, $-0.023\,\hbar$), as the model predicts for $\qpar\gg q^*$. Observing the saturation requires resolving the splitting $\kappa\eta$, which at the measured $q_z$ is only 0.35~meV per percent, well below current RIXS resolution (15--30~meV~\cite{Grimes2026}), and therefore RIXS probes the linear response of the group sum.

\textit{Prediction.} At the measured $\qpar$ the $d$-wave component of $H_{38}$ equals the $g$-wave one at $\eta^*=1.26\%$ for PBEsol when $b_4$ and $b_2$ are anchored to the nondiagonal values (1.6\% from the $3\times3\times3$ interpolation alone) [Fig.~\ref{fig:pred}(a)]; PBE+D3 lies about 11\% higher (1.4\%). Weighting each branch by its oxygen character or by its mode effective charge or its square, to which Ref.~\cite{Grimes2026} takes the RIXS intensity as proportional, changes $\eta^*$ by 5 and 1\%, respectively. Both values of $\eta^*$ lie within the strains reached in oxide membranes~\cite{Dong2019,Xu2020,Hong2020}. Directly at the measured wave vectors [Fig.~\ref{fig:strain}(d)], a tensile strain of $1\%$ reduces $H_{38}$ by 78\% at $\mathbf q_1$ and by 59 and 59\% at $\mathbf q_2$ and $\mathbf q_4$ (anchored values), and $H_{38}(\mathbf q_1)$ changes sign at $\eta\approx1.3\%$. $\mathbf q_3$ lies on a mirror that survives under $\eta$, so that $H_{38}(\mathbf q_3)=0$ remains a symmetry check. The cleanest signature is the dichroism along the diagonal [$\phi=45^\circ$ or $135^\circ$, Fig.~\ref{fig:pred}(c)]: zero by symmetry in the unstrained membrane, the group helicity reaches $0.023\,\hbar$ at $\eta=1\%$, 63\% of the maximum unstrained amplitude on the same ring (anchored), and with the 30~meV resolution of the switching experiment~\cite{Grimes2026} the 38~meV feature itself becomes dichroic in the partial spectrum of the converged branches, with a broadened helicity of about $-0.02\,\hbar$ at $\eta=1\%$ ($-0.040\,\hbar$ at $2\%$), followed by a weaker feature of opposite sign near 64~meV; at 15~meV resolution the two features separate. The branches below 26~meV add a tail at 30~meV resolution whose sign is not converged with supercell size, so that the 15~meV mode is preferable for this test. Shear strain instead breaks the mirror through $\mathbf q_3$, where $H_{38}=0.023\,\hbar$ at $\gamma=2\%$ ($2\times2\times2$, a lower bound given the supercell trend).

\section{Discussion}
In summary, we have shown that strain is a second, independent knob for phonon chirality in BaTiO$_3$, i.e., polarization reverses the sign of the texture, whereas strain changes its angular class, with the $d$-wave term guaranteed by symmetry and its magnitude fixed by the first-principles calculations. The headline quantity, $\eta^*$, is a ratio of symmetry-resolved amplitudes; it agrees within about 10\% between two functionals that differ in their tetragonality. Checked without interpolation on the measured ring and next to it, respectively, the $3\times3\times3$ supercell overestimates the unstrained amplitude $b_4$ by 24\% and underestimates the strained amplitude $b_2$ by 3\%; anchored to the nondiagonal calculations, $\eta^*=1.26\%$ for PBEsol, with an uncertainty of about 10\% from the single-point anchors. The convergence of $b_4$ with supercell size is not monotonic ($2\times2\times2$ lies closer to the exact value at this wave vector), which identifies the long-range part of the unstrained force constants as the main numerical uncertainty.

Our calculations keep the polarization along the membrane normal, as in the electrically switched membranes of Ref.~\cite{Grimes2026}. The strains considered are comparable to the bulk tetragonality ($c/a-1\approx0.75\%$ experimentally), and along the tensile axis the in-plane polar mode softens further (Table~\ref{tab:gamma}), so that a rotation of $\mathbf P$ towards the plane, as observed in bent membranes~\cite{Dong2019} and found in first-principles calculations for 1--2\% uniaxial tension in BaTiO$_3$~\cite{Qian2026}, may compete. On the other hand, an out-of-plane bias field stabilizes $\mathbf P\parallel\hat z$, while uniaxial compression, whose isotropic part favors an out-of-plane polarization, gives the same $|\eta|$ with the opposite sign of the $d$-wave term. A tilt of $\mathbf P$ would lower the symmetry further and add harmonics to the $d$-wave term without removing it. The ratio $\eta^*$ is also robust against the overestimated polarization of the generalized-gradient functionals, because $b_4$ and $b_2$ are both odd in $\mathbf P$.

Finally, our results are harmonic and refer to $T=0$, whereas the experiment was performed at room temperature (295~K). This sets the main limit of the present work. Below 26~meV the transverse acoustic branches hybridize with the partner of the soft polar mode, whose harmonic instability reflects the long-range, chain-like correlation of the Ti displacements in BaTiO$_3$~\cite{Ghosez1999}; their helicities are not converged even in $3\times3\times3$ supercells and change sign between supercells. At room temperature these modes are stabilized by anharmonicity, and the Ti displacements have a partial order--disorder character~\cite{Zalar2003}, so that their chirality is a finite-temperature property that calls for anharmonic methods, such as the self-consistent phonon and effective-potential approaches already applied to BaTiO$_3$~\cite{Ehsan2021} and to other anharmonic solids~\cite{Errea2014,Hellman2011}. Since strain softens the polar mode along the tensile axis (Table~\ref{tab:gamma}) and these branches respond most strongly in our harmonic calculations, the chirality of soft-mode-hybridized phonons near the ferroelectric instability, whose circular motion can be driven coherently in perovskites~\cite{Basini2024} and which can become chiral under strain in ferroelectric textures~\cite{Dey2025}, is an open question that we find particularly promising. The 38~meV group lies at least 6~meV above this manifold, and we expect finite-temperature renormalization of its $E$ doublets to change its amplitudes only quantitatively. We report helicities, which Grimes \emph{et al.}\ identified as the quantity governing the dichroic contrast~\cite{Grimes2026}; a calculation of the full cross section would make the comparison quantitative. The same mechanism should apply to other tetragonal ferroelectrics, e.g., PbTiO$_3$, and to films on anisotropic substrates, which might provide a built-in $d$-wave texture without an external stage. Together with the electric-field switching of Ref.~\cite{Grimes2026}, strain would therefore allow both the sign and the symmetry of phonon chirality to be set on the same membrane platform, and we hope that these predictions encourage such measurements.

\begin{acknowledgments}
E.S.M. acknowledges financial support from Proyecto Interno USM 2026 PI\_LIR\_26\_12. J.M.F. thanks Universidad Central de Chile.
\end{acknowledgments}

\textit{Data availability.} The data that support the findings of this article are available from the authors upon reasonable request.

\textit{AI disclosure.} The authors used Claude (Anthropic) to assist with workflow scripts for the calculations, data processing, figure preparation and language editing. The authors designed the study, checked all calculations, results and references, and take full responsibility for the content.

\bibliography{references}
\end{document}